\documentstyle[12pt]{article}

\catcode`\@=11

\let\DOTSI\relax
\def\RIfM@{\relax\ifmmode}
\def\FN@{\futurelet\next}
\newcount\intno@
\def\iint{\DOTSI\intno@\tw@\FN@\ints@}
\def\iiint{\DOTSI\intno@\thr@@\FN@\ints@}
\def\iiiint{\DOTSI\intno@4 \FN@\ints@}
\def\idotsint{\DOTSI\intno@\z@\FN@\ints@}
\def\ints@{\findlimits@\ints@@}
\newif\iflimtoken@
\newif\iflimits@
\def\findlimits@{\limtoken@true\ifx\next\limits\limits@true
 \else\ifx\next\nolimits\limits@false\else
 \limtoken@false\ifx\ilimits@\nolimits\limits@false\else
 \ifinner\limits@false\else\limits@true\fi\fi\fi\fi}
\def\multint@{\int\ifnum\intno@=\z@\intdots@                                
 \else\intkern@\fi                                                          
 \ifnum\intno@>\tw@\int\intkern@\fi                                         
 \ifnum\intno@>\thr@@\int\intkern@\fi                                       
 \int}                                                                      
\def\multintlimits@{\intop\ifnum\intno@=\z@\intdots@\else\intkern@\fi
 \ifnum\intno@>\tw@\intop\intkern@\fi
 \ifnum\intno@>\thr@@\intop\intkern@\fi\intop}
\def\intic@{\mathchoice{\hskip.5em}{\hskip.4em}{\hskip.4em}{\hskip.4em}}
\def\negintic@{\mathchoice
 {\hskip-.5em}{\hskip-.4em}{\hskip-.4em}{\hskip-.4em}}
\def\ints@@{\iflimtoken@                                                    
 \def\ints@@@{\iflimits@\negintic@\mathop{\intic@\multintlimits@}\limits    
  \else\multint@\nolimits\fi                                                
  \eat@}                                                                    
 \else                                                                      
 \def\ints@@@{\iflimits@\negintic@
  \mathop{\intic@\multintlimits@}\limits\else
  \multint@\nolimits\fi}\fi\ints@@@}
\def\intkern@{\mathchoice{\!\!\!}{\!\!}{\!\!}{\!\!}}
\def\plaincdots@{\mathinner{\cdotp\cdotp\cdotp}}
\def\intdots@{\mathchoice{\plaincdots@}
 {{\cdotp}\mkern1.5mu{\cdotp}\mkern1.5mu{\cdotp}}
 {{\cdotp}\mkern1mu{\cdotp}\mkern1mu{\cdotp}}
 {{\cdotp}\mkern1mu{\cdotp}\mkern1mu{\cdotp}}}

\newif\iffirstchoice@
\firstchoice@true
\def\textfonti{\the\textfont\@ne}
\def\textfontii{\the\textfont\tw@}
\def\text{\RIfM@\expandafter\text@\else\expandafter\text@@\fi}
\def\text@@#1{\leavevmode\hbox{#1}}
\def\text@#1{\mathchoice
 {\hbox{\everymath{\displaystyle}\def\textfonti{\the\textfont\@ne}%
  \def\textfontii{\the\textfont\tw@}\textdef@@ T#1}}
 {\hbox{\firstchoice@false
  \everymath{\textstyle}\def\textfonti{\the\textfont\@ne}%
  \def\textfontii{\the\textfont\tw@}\textdef@@ T#1}}
 {\hbox{\firstchoice@false
  \everymath{\scriptstyle}\def\textfonti{\the\scriptfont\@ne}%
  \def\textfontii{\the\scriptfont\tw@}\textdef@@ S\rm#1}}
 {\hbox{\firstchoice@false
  \everymath{\scriptscriptstyle}\def\textfonti
  {\the\scriptscriptfont\@ne}%
  \def\textfontii{\the\scriptscriptfont\tw@}\textdef@@ s\rm#1}}}
\def\textdef@@#1{\textdef@#1\rm\textdef@#1\bf\textdef@#1\sl\textdef@#1\it}
\def\DN@{\def\next@}
\def\eat@#1{}
\def\textdef@#1#2{%
 \DN@{\csname\expandafter\eat@\string#2fam\endcsname}%
 \if S#1\edef#2{\the\scriptfont\next@\relax}%
 \else\if s#1\edef#2{\the\scriptscriptfont\next@\relax}%
 \else\edef#2{\the\textfont\next@\relax}\fi\fi}

\def\Let@{\relax\iffalse{\fi\let\\=\cr\iffalse}\fi}
\def\vspace@{\def\vspace##1{\crcr\noalign{\vskip##1\relax}}}
\def\multilimits@{\bgroup\vspace@\Let@
 \baselineskip\fontdimen10 \scriptfont\tw@
 \advance\baselineskip\fontdimen12 \scriptfont\tw@
 \lineskip\thr@@\fontdimen8 \scriptfont\thr@@
 \lineskiplimit\lineskip
 \vbox\bgroup\ialign\bgroup\hfil$\m@th\scriptstyle{##}$\hfil\crcr}
\def\Sb{_\multilimits@}
\def\endSb{\crcr\egroup\egroup\egroup}
\def\Sp{^\multilimits@}

\newdimen\ex@
\def\rightarrowfill@#1{$#1\m@th\mathord-\mkern-6mu\cleaders
 \hbox{$#1\mkern-2mu\mathord-\mkern-2mu$}\hfill
 \mkern-6mu\mathord\rightarrow$}
\def\leftarrowfill@#1{$#1\m@th\mathord\leftarrow\mkern-6mu\cleaders
 \hbox{$#1\mkern-2mu\mathord-\mkern-2mu$}\hfill\mkern-6mu\mathord-$}
\def\leftrightarrowfill@#1{$#1\m@th\mathord\leftarrow\mkern-6mu\cleaders
 \hbox{$#1\mkern-2mu\mathord-\mkern-2mu$}\hfill
 \mkern-6mu\mathord\rightarrow$}
\def\overrightarrow{\mathpalette\overrightarrow@}
\def\overrightarrow@#1#2{\vbox{\ialign{##\crcr\rightarrowfill@#1\crcr
 \noalign{\kern-\ex@\nointerlineskip}$\m@th\hfil#1#2\hfil$\crcr}}}

\def\overleftarrow{\mathpalette\overleftarrow@}
\def\overleftarrow@#1#2{\vbox{\ialign{##\crcr\leftarrowfill@#1\crcr
 \noalign{\kern-\ex@\nointerlineskip}$\m@th\hfil#1#2\hfil$\crcr}}}
\def\overleftrightarrow{\mathpalette\overleftrightarrow@}
\def\overleftrightarrow@#1#2{\vbox{\ialign{##\crcr\leftrightarrowfill@#1\crcr
 \noalign{\kern-\ex@\nointerlineskip}$\m@th\hfil#1#2\hfil$\crcr}}}
\def\underrightarrow{\mathpalette\underrightarrow@}
\def\underrightarrow@#1#2{\vtop{\ialign{##\crcr$\m@th\hfil#1#2\hfil$\crcr
 \noalign{\nointerlineskip}\rightarrowfill@#1\crcr}}}

\def\underleftarrow{\mathpalette\underleftarrow@}
\def\underleftarrow@#1#2{\vtop{\ialign{##\crcr$\m@th\hfil#1#2\hfil$\crcr
 \noalign{\nointerlineskip}\leftarrowfill@#1\crcr}}}
\def\underleftrightarrow{\mathpalette\underleftrightarrow@}
\def\underleftrightarrow@#1#2{\vtop{\ialign{##\crcr$\m@th\hfil#1#2\hfil$\crcr
 \noalign{\nointerlineskip}\leftrightarrowfill@#1\crcr}}}

\catcode`\@=\active

\def\frac#1#2{{#1 \over #2}}

\def\stackunder#1#2{\mathrel{\mathop{#2}\limits_{#1}}}

\newcount\GRAPHICSTYPE
\def\GRAPHICSPS#1{%
\ifnum\GRAPHICSTYPE=1 language "PS", include "#1"\else%
ps: #1\fi}

\def\graffile#1#2#3#4{\leavevmode\raise -#4 \hbox{%
\raise #3 \hbox{\rule{0.003in}{0.003in}\special{#1}}}%
{\raise -#4 \hbox to #2 {\vrule height#3 width0in depth0in\hfil}}%
}

\def\draftbox#1#2#3#4{\leavevmode\raise -#4 \hbox{\frame{\rlap{\protect\tiny
#1}%
\hbox to #2{\vrule height#3 width0in depth0in\hfil}}}}

\newcount\draft
\def\GRAPHIC#1#2#3#4#5{\ifnum\draft=1 \draftbox{#2}{#3}{#4}{#5}\else%
\graffile{#1}{#3}{#4}{#5}\fi}

\def\addtoLaTeXparams#1{\edef\LaTeXparams{\LaTeXparams #1}}

\def\doFRAMEparams#1{\readFRAMEparams#1\end}
\def\readFRAMEparams#1{%
\ifx#1\end%
\let\next=\relax%
\else%
\ifx#1i%
\dispkind=0%
\fi%
\ifx#1d%
\dispkind=1%
\fi%
\ifx#1f%
\dispkind=2%
\fi%
\ifx#1t%
\addtoLaTeXparams{t}%
\fi%
\ifx#1b%
\addtoLaTeXparams{b}%
\fi%
\ifx#1p%
\addtoLaTeXparams{p}%
\fi%
\ifx#1h%
\addtoLaTeXparams{h}%
\fi%
\let\next=\readFRAMEparams%
\fi%
\next%
}

\def\IFRAME#1#2#3#4#5{\GRAPHIC{#5}{#4}{#1}{#2}{#3}}

\def\DFRAME#1#2#3#4{
  \begin{center}
    \GRAPHIC{#4}{#3}{#1}{#2}{0in}
  \end{center}
}

\def\FFRAME#1#2#3#4#5#6#7{
  \begin{figure}[#1]
    \begin{center}
      \GRAPHIC{#7}{#6}{#2}{#3}{0in}
    \end{center}
    \caption{\label{#5}#4}
  \end{figure}
}

\def\FRAME#1#2#3#4#5#6#7#8{%
\newcount\dispkind%
\def\LaTeXparams{}%
\dispkind=0%
\def\LaTeXparams{}%
\doFRAMEparams{#1}%
\ifnum\dispkind=0%
\IFRAME{#2}{#3}{#4}{#7}{#8}%
\else
  \ifnum\dispkind=1
    \DFRAME{#2}{#3}{#7}{#8}
  \else
    \ifnum\dispkind=2
      \FFRAME{\LaTeXparams}{#2}{#3}{#5}{#6}{#7}{#8}
    \fi
  \fi
\fi
}

\catcode`\@=11

\long\def\QQQ#1#2{}
\def\QTP#1{}
\long\def\QQA#1#2{}

\def\EXPAND#1[#2]#3{}
\def\NOEXPAND#1[#2]#3{}

\def\LaTeXparent#1{}

\@ifundefined{INDEX}{\def\INDEX#1#2{}{}}{}
\@ifundefined{SUBINDEX}{\def\SUBINDEX#1#2#3{}{}{}}{}
\def\initial#1{\bigbreak{\raggedright\large\bf #1}\kern 2pt\penalty3000}

\@ifundefined{abstract}{%
\def\abstract{\if@twocolumn
\section*{Abstract (Not appropriate in this style!)}
\else \small
\begin{center}
{\bf Abstract\vspace{-.5em}\vspace{0pt}}
\end{center}
\quotation
\fi}}{}
\@ifundefined{endabstract}{%
\def\endabstract{\if@twocolumn\else\endquotation\fi}}{}
\@ifundefined{maketitle}{\def\maketitle#1{}}{}
\@ifundefined{affiliation}{\def\affiliation#1{}}{}
\@ifundefined{proof}{}{}
\@ifundefined{newfield}{\def\newfield#1#2{}}{}
\@ifundefined{chapter}{\def\chapter#1{\par(Chapter head:)#1\par }}{}
\@ifundefined{part}{\def\part#1{\par(Part head:)#1\par }}{}
\@ifundefined{section}{\def\part#1{\par(Section head:)#1\par }}{}
\@ifundefined{subsection}{\def\part#1{\par(Subsection head:)#1\par }}{}
\@ifundefined{subsubsection}{\def\part#1{\par(Subsubsection head:)#1\par }}{}
\@ifundefined{paragraph}{\def\part#1{\par(Subsubsubsection head:)#1\par }}{}
\@ifundefined{subparagraph}{\def\part#1{\par(Subsubsubsubsection head:)#1\par
}}{}

\newdimen\theight
\def \Column{%
             \vadjust{\setbox0=\hbox{\scriptsize\quad\quad tcol}%
             \theight=\ht0
             \advance\theight by \dp0    \advance\theight by \lineskip
             \kern -\theight \vbox to \theight{\rightline{\rlap{\box0}}%
             \vss}%
             }}%

\def\qed{\ifhmode\unskip\nobreak\fi\ifmmode\ifinner\else\hskip5\p@\fi\fi
 \hbox{\hskip5\p@\vrule width4\p@ height6\p@ depth1.5\p@\hskip\p@}}
\catcode`@=12 

\makeatletter

\newtheorem{theorem}{Theorem}[section]

\author{Alexey Sevostyanov\\ 
Steklov Mathematical Institute, St. Petersburg
}

\title{The classical r-matrix method for nonlinear sigma-model
}
\QQQ{Language}{
American English
}

\begin{document}

\maketitle
\begin{abstract}
The canonical Poisson structure of nonlinear sigma-model is presented as a
Lie-Poisson r-matrix bracket on coadjoint orbits. It is shown that the Poisson
structure of this model is determined by some `hidden singularities' of the
Lax matrix.
\end{abstract}

\section*{Introduction}

Two-dimensional nonlinear sigma models were studied for almost 20 years.
Their full treatment in quantum case proved to be much more difficult than
for other related relativistic models, e.g. the Sine Gordon equation.
Surprisingly, a consistent r-matrix formulation of these models (which is a
neccessary prerequisite of the study of the quantum case) was lacking,
although a Lax pair for chiral model was found by Zakharov and Mikhailov
many years ago. The purpose of this note is to explain the origin of the
Zakharov-Mikhailov Lax pair and its generalizations and of the associated
Poisson structures in the r-matrix language. As it appears, the relevant
r-matrices are non-unitary; they belong to the hierarchy associated with the
standard rational r-matrix. Our method also applies to nonlinear sigma
models with values in a riemannian symmetric space; it represents a first
step towards a solution of the corresponding quantum problem. (Although the
particle spectrum and the corresponding factorized scattering matrices have
been guessed many years ago by Zamolodchikov and later Faddeev and
Reshetikhin \cite{fr} were able to reproduce this result using an infinite
spin limit in an appropriate lattice model, a systematic treatment of chiral
models on the lines of the Quantum Inverse Scattering Method still seems to
be lacking.) It should be noted that the non-unitarity of the r-matrix
poses additional problems in our approach. The crucial point is to find
consistent Poisson bracket relations for the monodromy matrix of the Lax
operator. The anomalies in these relations are directly related to the
non-unitarity of the r-matrix, and although their regualrization in some
cases is possible, for chiral models this technique fails.

In this paper we shall discuss sigma models with values in a semisimple Lie
group (following a suggestion of L.D.Faddeev we shall call them principal
chiral models), as well as sigma models with values in Riemannian symmetric
spaces.

I am grateful to L.D.Faddeev, F.A.Smirnov, and A.Yu.Alekseev for helpful
discussions.

I would like to thank M.A.Semenov-Tian-Shansky who pointed out to me the
elegant ad\`elic construction of the rational r-matrices and of the related
linear hierarchies of Poisson brackets \cite{ch}, \cite{rs2}.

\section{ Hamiltonian formulation of chiral fields.}

We remind some standard facts on the canonical Poisson structures on
cotangent bundles of Lie groups \cite{ft}. Let ${\em G}$ be a Lie group, $%
{\bf g}$ its Lie algebra, and ${\bf g}^{*}$ the dual space of ${\bf g}$. The
cotangent bundle $T^{*}{\em G}$ admits two canonical trivializations by left
and right translations, respectively. If $L_g:h\longmapsto gh$ is a left
translation on $\widehat{{\em G}},$ then

$$
dL_g^{*}:T_g^{*}{\em G}\rightarrow T_e^{*}{\em G}\simeq {\bf g}^{*}.
$$

Similarly, if $R_g:h\longmapsto hg$ is a right translation, then

$$
dR_g^{*}:T_g^{*}{\em G}\rightarrow T_e^{*}{\em G}\simeq {\bf g}^{*}.
$$

Thus if $\left( g,\xi _g\right) \in T^{*}{\em G},\xi _g\in T_g^{*}{\em G},$
then

\begin{equation}
\label{1.1}
\begin{array}{c}
\left( g,\xi _g\right)
\stackrel{L}{\mapsto }\left( g,l_t\right) ,l_t=dL_g^{*}\xi _g, \\ \left(
g,\xi _g\right)
\stackrel{R}{\mapsto }\left( g,r_t\right) ,r_t=-dR_g^{*}\xi _g, \\ T^{*}{\em %
G}\stackrel{L,R}{\rightarrow }{\em G}\times {\bf g}^{*}
\end{array}
\end{equation}
are the left and right trivializations. Clearly, we have
\begin{equation}
\label{1.2}r_t=-Ad^{*}g\left( l_t\right) .
\end{equation}
The minus sign in (\ref{1.1}), (\ref{1.2}) reflects the fact that the Lie
algebras of left- and right-invariant vector fields on $G$ are
anti-isomorphic. Let $\widehat{{\bf g}}=C^\infty \left( S^1,{\bf g}\right) $
be the current algebra and $\widehat{G}$ the corresponding current group.
Using the trivializations of $T^{*}G$ described above the cotangent bundle $%
T^{*}$ $\widehat{G}$ may be identified with $\widehat{{\em G}}\times
\widehat{{\bf g}}^{*}.$ The canonical Poisson bracket on $T^{*}$ $\widehat{G}
$ \cite{a}, \cite{kr} is described as follows:

Every functional on $T^{*}$ $\widehat{G}$ may be represented as a function
of two variables $g\in \widehat{G}$ and $l_t\in \widehat{{\bf g}}^{*}$ (or $%
r_t\in \widehat{{\bf g}}^{*},$ depending on the trivialization chosen) . The
derivatives with respect to these variables are defined by:

\begin{equation}
\label{1.3}
\begin{array}{c}
\left\langle
{\rm D}^{\prime }\varphi \left( g,l_t\right) ,X\right\rangle =\frac
d{dt}\mid _{t=0}\varphi \left( ge^{tX},l_t\right) , \\ \left\langle
X_\varphi \left( g,l_t\right) ,Y\right\rangle =\frac d{dt}\mid _{t=0}\varphi
\left( g,l_t+tY\right) , \\
\varphi \in Fun\left( T^{*}\widehat{G}\right) ,{\rm D}^{\prime }\varphi
,Y\in \widehat{{\bf g}}^{*};X_\varphi ,X\in \widehat{{\bf g}},
\end{array}
\end{equation}

where $\left\langle \cdot ,\cdot \right\rangle $ is the natural pairing
between $\widehat{{\bf g}}$ and $\widehat{{\bf g}}^{*}$ . In our case we may
identify $\widehat{{\bf g}}$ and $\widehat{{\bf g}}^{*}$ using an invariant
scalar product on $\widehat{{\bf g}}$ :

\begin{equation}
\label{1.4}\left\langle X\left( x\right) ,Y\left( x\right) \right\rangle
=\int\limits_0^{2\pi }tr\left( X\left( x\right) Y\left( x\right) \right) dx,
\end{equation}

where $tr$ is an invariant bilinear form on ${\bf g}$ .

Now we are ready to define the canonical Poisson bracket on $T^{*}\widehat{G}
$ :

\begin{equation}
\label{1.5}\left\{ \varphi ,\psi \right\} \left( g,l_t\right) =\left\langle
{\rm D}^{\prime }\psi ,X_\varphi \right\rangle -\left\langle {\rm D}^{\prime
}\varphi ,X_\psi \right\rangle +\left\langle l_t,\left[ X_\varphi ,X_\psi
\right] \right\rangle .
\end{equation}

In tensor notations we have the following formulas for the brackets of
matrix elements of $g$ and $l_t$ :

\begin{equation}
\label{1.6}
\begin{array}{c}
\left\{ l_t\left( x\right)_1 ,l_t\left( y\right)_2 \right\} =\frac 12\left[
t,l_t\left( x\right)_1 -l_t\left( y\right)_2 \right] \delta \left(
x-y\right) , \\
\left\{ g_1\left( x\right) ,l_t\left( y\right)_2 \right\} =-g_1\left(
x\right) t\delta \left( x-y\right) , \\
\left\{ g_1\left( x\right) ,g_2\left( y\right) \right\} =0,
\end{array}
\end{equation}

where $l_t\left( x\right) _1=l_t\left( x\right) \otimes I,l_t\left( x\right)
_2=I\otimes l_t\left( x\right) $ , and $t$ is the tensor Casimir in ${\bf g}%
\otimes {\bf g}$ .

Our sign convention in formula (\ref{1.1}) makes the mapping (\ref{1.2})
which converts left trivialization into right trivialization a Poisson
mapping.

The following series of results is connected with the Hamiltonian reduction
on $T^{*}\widehat{G}$ \cite{a}, \cite{am}. The natural actions of $\widehat{G%
}$ on itself by left and right translations may be canonically lifted to $%
T^{*}\widehat{G}$; if we use the trivialization of $T^{*}\widehat{G}$ by
left translations the lifted actions are given by

\begin{equation}
\label{1.7}
\begin{array}{c}
\widehat{G}\times T^{*}\widehat{G}\stackrel{L,R}{\rightarrow }T^{*}\widehat{G%
}, \\ L_h\left( g,l_t\right) =\left( hg,l_t\right) , \\
R_h\left( g,l_t\right) =\left( gh,Ad^{*}h^{-1}\left( l_t\right) \right)
\end{array}
\end{equation}

These actions are Hamiltonian; the Hamiltonians which correspond to $X\in
\widehat{{\bf g}}$ are given by

\begin{equation}
\label{1.8}
\begin{array}{c}
H_X^L\left( g,l_t\right) =-\left\langle X,Ad^{*}g\left( l_t\right)
\right\rangle , \\
H_H^R=\left\langle X,l_t\right\rangle , \\
X\in \widehat{{\bf g}}.
\end{array}
\end{equation}

The corresponding moment maps are

\begin{equation}
\label{1.10}
\begin{array}{c}
\mu _L=-Ad^{*}g\left( l_t\right) =r_t, \\
\mu _R\left( g,l_t\right) =l_t.
\end{array}
\end{equation}

Let $\theta _L,\theta _{R\text{ }}$ be the left-invariant (respectively, the
right-invariant) Maurer-Cartan form on $G.$ The left-invariant
(right-invariant) current associated with $g\in \hat G$ is defined by
$$
l_xdx=g^{*}\theta _L,r_xdx=-g^{*}\theta _R
$$
If $G$ is a matrix group we have simply%
$$
l_x(x)=g^{-1}\partial _xg(x),r_x(x)=-\partial _xg(x)g^{-1}.
$$

{\it Remark}. The notation we use suggests that $(l_t,l_x)$ and $(r_t,r_x)$
are two components of a single two-dimensional Noether current
(left-invariant or right-invariant, respectively). This notation will be
motivated later.

The currents $l_x,r_x$ parametrize the quotient space obtained by reduction
of $T^{*}\widehat{G}$ over the action of the subgroup $G$ of constant loops
by left (respectively, right) translations. The moment map which coresponds
to this action is given by

\begin{equation}
\label{1.11}\mu _L^{{\em G}}\left( g,l_t\right) =\int\limits_0^{2\pi
}r_t\left( x\right) dx.
\end{equation}

The quotient space ${\em G}\setminus T^{*}\widehat{G}$ may be identified
with $\widehat{{\bf g}}\times \widehat{{\bf g}}^{*}$ via the mapping:
$$
\left( g,l_t\right) \mapsto \left( l_x,l_t\right)
$$
The coordinates $l_x,l_t$ have the following Poisson brackets in the tensor
notations:

\begin{equation}
\label{1.12}
\begin{array}{c}
\left\{ l_t\left( x\right) _1,l_x\left( y\right) _2\right\} =\frac 12\left[
t,l_x\left( x\right) _1-l_x\left( y\right) _2\right] \delta \left(
x-y\right) - \\
-t\delta ^{\prime }\left( x-y\right) , \\
\left\{ l_t\left( x\right) _1,l_t\left( y\right) _2\right\} =\frac 12\left[
t,l_t\left( x\right) _1-l_t\left( y\right) _2\right] \delta \left(
x-y\right) , \\
\left\{ l_x\left( x\right) _1,l_x\left( y\right) _2\right\} =0.
\end{array}
\end{equation}

This Poisson structure is slightly degenerate and the space ${\em G}%
\setminus T^{*}\widehat{G}$ decomposes into symplectic leaves; among these
latter there is one which corresponds to the zero moment. This leaf is the
reduced phase space over the point $\mu _L^{{\em G}}\left( g,l_t\right) =0$
\cite{a} ,\cite{am} .We denote this symplectic manifold by ${\cal M}$.

Now we define the principal chiral field as a Hamiltonian system on this
symplectic manifold with the Hamiltonian function:

\begin{equation}
\label{1.14}H=\frac 12\int\limits_0^{2\pi }tr\left( l_xl_x+l_tl_t\right) dx.
\end{equation}

The equations of motion have the form:

\begin{equation}
\label{1.15}
\begin{array}{c}
\partial _tl_\mu =\left\{ H,l_\mu \right\} , \\
\partial _xl_x=\partial _tl_t, \\
\partial _xl_t-\partial _tl_x+\left[ l_x,l_t\right] =0.
\end{array}
\end{equation}

The last equation is the zero curvature condition which serves to restore
the group variable $g\left( x\right) \in \widehat{G}$ using the variables $%
l_x,l_t$ modulo the left action of constant loops.

To justify the notation introduced above let us observe that the action
functional which corresponds to our choice of the Hamiltonian is given by

\begin{equation}
\label{1.13}S\left( g\right) =\frac 12\int tr\left( l_xl_x-l_tl_t\right)
dxdt.
\end{equation}
where $l_t=g^{-1}\partial _tg,l_x=g^{-1}\partial _xg;$ clearly, the Legendre
transform associated with (\ref{1.13}) identifies $l_t,l_x$ with our
canonical variables and hence our notation is consistent. The same remark
applies, of course, to right currents $r_t=-\partial _tgg^{-1},r_x=-\partial
_xgg^{-1},$ since both the Hamiltonian (\ref{1.14}) and the action
functional (\ref{1.13}) are Ad-invariant.

We turn to the study of chiral fields with values in symmetric spaces. Our
exposition follows \cite{fs}; we add some details on the canonical formalism.

Let  ${\bf g}={\bf k}\stackrel{.}{+}{\bf p}$ be a Cartan decomposition \cite
{h}, i.e.

\begin{equation}
\label{1.16}\left[ {\bf k},{\bf k}\right] \subset {\bf k},\left[ {\bf k},%
{\bf p}\right] \subset {\bf p},\left[ {\bf p},{\bf p}\right] \subset {\bf k}%
,
\end{equation}

so that ${\bf k}$ is a Lie subalgebra in ${\bf g}$ . Put $\widehat{{\bf k}}%
=C^\infty \left( S^1,{\bf k}\right) $ , and let $\widehat{K}$ be the
corresponding subgroup in $\widehat{G}$ . There is a Hamiltonian action of $%
\widehat{K}$ on $T^{*}\widehat{G}$ which arises from right translations by
elements of $\widehat{G}$ . Its moment map is given by

\begin{equation}
\label{1.17}\mu _R^{{\em K}}\left( g,l_t\right) =P_{{\bf k}}l_t,
\end{equation}

where $P_{{\bf k}}$ is the orthogonal projection operator onto the subspace $%
{\bf k}$ $\subset {\bf g}$ .

Since the left and right actions of $\widehat{{\em G}}$ on $T^{*}\widehat{%
{\em G}}$ commute, the above defined action of $\widehat{{\em K}}$ on $T^{*}%
\widehat{{\em G}}$ generates the action of $\widehat{{\em K}}$ on ${\em G}%
\backslash T^{*}\widehat{{\em G}}$ and even on the symplectic submanifold $%
{\cal M}$ in this space with the same moment map (\ref{1.17}).We introduce
the quotient space over this action ${\em G}\backslash T^{*}\widehat{{\em G}}%
/\widehat{K}$. The quotient Poisson structure on this space is degenerate.
The phase space of the chiral field is the symplectic leaf corresponding to
the zero values of the left and right moments of the actions of ${\em G}$
and $\widehat{{\em K}}$ $\mu _L^{{\em G}}\left( g,l_t\right) =\mu _R^{{\em K}%
}\left( g,l_t\right) =0$ . This is the reduced phase space over these
actions. We denote it by ${\cal M}_1$.

Let us choose the following coordinates on the space ${\em G}\backslash
T^{*} \widehat{{\em G}}$ :

\begin{equation}
\label{1.18}P_{{\bf k}}l_\mu =A_\mu ,P{\bf _p}l_\mu =B_\mu ,\mu =t,x.
\end{equation}

Now we describe the Poisson structure of the phase space. First of all we
calculate the Poisson brackets of the variables $A_x,B_x,B_t$ .

Let $t_{A=}\left( P_{{\bf k}}\otimes P_{{\bf k}}\right) t,t_{B=}\left( P_{%
{\bf p}}\otimes P_{{\bf p}}\right) t$ be the ${\bf k}$ and ${\bf p}$
components of the Casimir element of ${\bf g}$ ,. $t=t_A+t_B$. In this
realization we have the following Poisson brackets for $A_x$ and $B_\mu $ :

\begin{equation}
\label{1.19}
\begin{array}{c}
\left\{ B_t\left( x\right) _1,B_t\left( y\right) _2\right\} =0, \\
\left\{ B_t\left( x\right) _1,B_x\left( y\right) _2\right\} =\frac 12\left[
t_B,A_x\left( x\right) _1-A_x\left( y\right) _2\right] \delta \left(
x-y\right) - \\
-t_B\delta ^{\prime }\left( x-y\right) , \\
\left\{ B_t\left( x\right) _1,A_x\left( y\right) _2\right\} =\left[
t_A,B_x\left( x\right) _1\right] \delta \left( x-y\right) .
\end{array}
\end{equation}

These brackets give the Poisson structure on the subspace in ${\em G}%
\backslash T^{*}\widehat{{\em G}}$ (and hence on ${\cal M}$ ) on which $\mu
_R^{{\em K}}\left( g,l_t\right) =0$ .The space ${\cal M}_1$ carries the
quotient Poisson structure of the Poisson structure (\ref{1.19}) under the
right action of the group $\widehat{{\em K}}$ on ${\cal M}$ .

The chiral field is a Hamiltonian system on the   phase space  ${\cal M}_1$%
described above with the Hamiltonian function:

\begin{equation}
\label{1.22}H=\frac 12\int\limits_0^{2\pi }tr\left( B_xB_x+B_tB_t\right) dx,
\end{equation}

This function is invariant under the right action $\widehat{{\em K}}$ on $%
{\cal M}$ so that it is a well defined function on the phase space. The
equations of motion have the form:

\begin{equation}
\label{1.21}
\begin{array}{c}
\partial _tB_t=\partial _xB_x+\left[ A_x,B_x\right] , \\
-\partial _tA_x+\left[ B_x,B_t\right] =0, \\
\partial _xB_t-\partial _tB_x+\left[ A_x,B_t\right] =0.
\end{array}
\end{equation}

As above, the two last equations are zero curvature conditions which serve
to restore the field variable $g\left( x\right) \in \widehat{G}/\widehat{K}$
given the currents $A_x$ and $B_\mu $ (this correspondence is unique modulo
constant loops).

Note that the functions $A,B$  are  coordinate functions on the space ${\em G%
}\backslash T^{*}\widehat{{\em G}}$ which is larger than the phase space $%
{\cal M}_1$ ; the genuine observables for the reduced system are gauge
invariant functionals of $A,B.$ However it is convenient to look at the
evolution on this larger space as well; as we shall see it admits a Lax
representation.

As well as for the principal chiral field one can calculate the Lagrangian
function given the Hamiltonian function and the Poisson structure of the
phase space. This calculations gives the well known result:

\begin{equation}
\label{1.20}S\left( g\right) =\frac 12\int tr\left( B_xB_x-B_tB_t\right)
dxdt.
\end{equation}

This shows that $A_x,B_x,B_t$ are the components of the current and $S$ is a
well defined function on $\widehat{{\em G}}/\widehat{{\em K}}$ .

If ${\bf g}=su(2)$ and ${\bf k}={\bf h}$ is the Cartan subalgebra of $su(2)$
our description of the nonlinear sigma-model is equivalent to the
formulation given in \cite{ft}. This formulation uses the space $T^{*}
\widehat{G}/\widehat{K}$ as the phase space of our model. In this
realization the coordinates on $T^{*}\widehat{G}/\widehat{K}$ are:

\begin{equation}
\label{1.23}
\begin{array}{c}
n=gXg^{-1},\pi =-\left[ r_t,n\right] , \\
\text{here }X=\left(
\begin{array}{cc}
i & 0 \\
0 & -i
\end{array}
\right) ,
\end{array}
\end{equation}

and it is supposed that $\mu _R^K\left( g,l\right) =P_{{\bf k}}l_t=0$ . We
don`t need the explicit form of the equations of motion in this realization
nor the formulas expressing the Poisson structure in terms of $n$ and $\pi $
. The reader may easily restore these formulas using \cite{ft}. For
instance, we rewrite the action (\ref{1.20}) in the terms of $n$ and $\pi $ :

\begin{equation}
\label{1.24}S=\frac 12\int \left( \left( \partial _xn\right) ^2-\left( \pi
\right) ^2\right) dxdt.
\end{equation}

\section{The coadjoint orbits formulation for principal chiral fields.}

In this section we shall develope the Lie-algebraic point of view on the
Poisson structure of the principal chiral field. Our goal is to propose an
r-matrix formulation of the Zakharov-Mikhailov Lax pair \cite{ft}, \cite{s1}%
. The Lax matrix of Zakharov and Mikhailov is a rational function on $CP_1$
with two poles at $\lambda =\pm 1$; there is also a 'hidden' singularity at $%
\lambda =0$ .  As we shall see, it is this latter singularity that
determines the Poisson structure of the chiral model. By contrast, apparent
singularities at $\lambda =\pm 1$ do not influence the Poisson structure;
instead each of them produces a series of local conservation laws. A
convenient formalism allowing to work with Lax matrices with arbitrary poles
uses the algebra of ad\`eles of rational functions; the rational r-matrix is
associated with the canonical decomposition of this algebra into
complementary subalgebras.

We remind some definitions \cite{rs2}. Let ${\bf g}$ is a semisimple Lie
algebra, ${\bf g}_{-}$ the Lie algebra of rational functions on $\overline{%
{\bf C}}$ with values in ${\bf g}.$ Let $\lambda _\nu $ be the local
parameter at $\nu \in \overline{{\bf C}},\lambda _\nu =\lambda -\nu ,\nu \in
{\bf C},\lambda _\infty =-\frac 1\lambda .$ Let ${\bf g}_\nu ={\bf g}\left(
\left( \lambda _\nu \right) \right) $ be the algebra of formal Laurent
series with values in ${\bf g}$ . Put ${\bf g}_{{\cal A}}=\coprod\limits_{%
\nu \in \overline{{\bf C}}}{\bf g}_\nu $ . In this formula it is supposed
that the sum is taken over the extended complex plane $\overline{{\bf C}}$
and for every element of ${\bf g}_{{\cal A}}$ only a finite number of series
in the direct sum are Laurent series and all the others are Taylor series.
The algebra ${\bf g}_{-}$ of rational functions with values in ${\bf g}$ is
embedded in ${\bf g}_{{\cal A}\text{ }}$ in the following way:

\begin{equation}
\label{2.1}X\left( \lambda \right) \rightarrow \bigoplus\limits_{\nu \in
\overline{{\bf C}}}X\left( \lambda \right) _\nu ;
\end{equation}

here $X\left( \lambda \right) \in {\bf g}_{-}$ and $X\left( \lambda \right)
_\nu $ is the Laurent series of $X\left( \lambda \right) $ at the point $\nu
$ . The subalgebra ${\bf g}_{-}$ is isotropic in ${\bf g}_{{\cal A}}$ with
respect to the invariant scalar product:

\begin{equation}
\label{2.2}
\begin{array}{c}
\left( X\left( \lambda \right) ,Y\left( \lambda \right) \right)
=\sum\limits_{\nu \in
\overline{{\bf C}}}Res_\nu \text{ }trX_\nu \left( \lambda _\nu \right) Y_\nu
\left( \lambda _\nu \right) d\lambda , \\ X\left( \lambda \right)
=\bigoplus\limits_{\nu \in \overline{{\bf C}}}X_\nu \left( \lambda _\nu
\right) ,Y\left( \lambda \right) =\bigoplus\limits_{\nu \in {\bf \overline{C}%
}}Y_\nu \left( \lambda _\nu \right) .
\end{array}
\end{equation}

For every element $X\in {\bf g}_{{\cal A}}$ there exists a rational function
$P_{-}X\left( \lambda \right) $ such that the principal parts of its Laurent
series of at all points coincide with the principal parts of the Laurent
series of $X_\nu \left( \lambda _\nu \right) $ at the same points. This
function is given by the formula:

\begin{equation}
\label{2.3}P_{-}X\left( \lambda \right) =\sum\limits_{\nu \in \overline{{\bf %
C}}}\stackunder{\nu }{Res}\text{ }tr\frac t{\lambda -\mu }X_\nu \left( \mu
_\nu \right) d\mu .
\end{equation}

So there exists a direct decomposition of the linear space ${\bf g}_{{\cal A}%
}$ :${\bf g}_{{\cal A}}={\bf g}_{-}\stackrel{.}{+}{\bf g}_{+}$ , where ${\bf %
g}_{-}$ is the algebra of rational functions and ${\bf g}_{+}$ is the
subalgebra of ${\bf g}_{{\cal A}}$ which consists of Taylor series.
Evidently, the subalgebra ${\bf g}_{+}$ is isotropic with respect to the
scalar product (\ref{2.2}) , so that there are natural pairings: ${\bf g}%
_{+}^{*}\simeq {\bf g}_{-},{\bf g}_{-}^{*}\simeq {\bf g}_{+}$ . And if we
define the projection operator $P_{-}$ onto $g(\lambda )$ by the formula (%
\ref{2.3}) and the complementary projection operator $P_{+}=I-P_{-}$ , then $%
P_{+}^{*}=P_{-}$ with respect to the scalar product (\ref{2.2}). The
rational r-matrix is defined by the standard formula \cite{rs2}, \cite{ft}:

\begin{equation}
\label{2.4}
\begin{array}{c}
r=P_{+}-P_{-}, \\
r^{*}=-r;
\end{array}
\end{equation}

It satisfies the modified classical Yang-Baxter equation. In standard
applications the space ${\bf g}_{-}$ is used as a model of the dual space $%
{\bf g}_{+}^{*};$ a generic point $L\in {\bf g}_{-}$ is regarded as a Lax
matrix and the Lie-Poisson bracket of ${\bf g}_{+}$ provides a 'universal'
hamiltonian structure for Lax equations in question. For our purposes we
modify this construction in the following way. To identify the space ${\bf g}%
_{{\cal A}}$ with ${\bf g}_{{\cal A}}^{*}$ we use the modified scalar
product on ${\bf g}_{{\cal A}}$ :

\begin{equation}
\label{2.5}\left( X\left( \lambda \right) ,Y\left( \lambda \right) \right)
_\phi =\sum\limits_{\nu \in \overline{{\bf C}}}\stackunder{\nu }{Res}\text{ }%
trX_\nu \left( \lambda _\nu \right) Y_\nu \left( \lambda _\nu \right) \phi
\left( \lambda \right) d\lambda ,
\end{equation}

where $\phi \left( \lambda \right) $ is some rational function with
numerical values. This scalar product allows to consider a nontrivial model
for the space ${\bf g}_{+}^{*}$ . Namely the space ${\bf g}_{+}$ is not
isotropic with respect to this scalar product and

\begin{equation}
\label{2.5.1}{\bf g}_{+}^{*}\simeq \left\{ \phi ^{-1}\left( \lambda \right)
X\left( \lambda \right) ,X\left( \lambda \right) \in {\bf g}_{-}\right\} .
\end{equation}

The rational r-matrix is not skew-symmetric with respect to this scalar
product:

\begin{equation}
\label{2.6}r^{*}=-\phi ^{-1}r\phi ,
\end{equation}

where $\phi $ denotes the operator of multiplication by $\phi $ .We use the
scalar product (\ref{2.5}) and the r-matrix (\ref{2.4}) to define the
r-matrix Lie-Poisson bracket for the central extension of the current
algebra $\widehat{{\bf g}}_{{\cal A}}=C^\infty \left( S^1,{\bf g}_{{\cal A}%
}\right) $ . It has the form \cite{ft},\cite{s2},\cite{rs1}:

\begin{equation}
\label{2.7a}
\begin{array}{c}
\left\{ \varphi ,\psi \right\} \left( L\right) =\int\limits_0^{2\pi
}dy\left( L,\left[ rX_\varphi ,X_\psi \right] +\left[ X_\varphi ,rX_\psi
\right] \right) _\phi - \\
-\int\limits_0^{2\pi }dy\left( \left( r+r^{*}\right) \partial _yX_\varphi
,X_\psi \right) _\phi , \\
L\in
\widehat{{\bf g}}_{{\cal A}}^{*},\text{and }X_\varphi \text{ is a derivative
of }\varphi : \\ \int\limits_0^{2\pi }dy\left( X_\varphi ,Y\right) _\phi
=\frac d{dt}\mid _{t=0}\varphi \left( L+tY\right) ,Y\in \widehat{{\bf g}}_{%
{\cal A}}^{*}.
\end{array}
\end{equation}

The Jacobi identity for this bracket follows from the Yang-Baxter equation
for $r$ . It is well known that this bracket may be restricted to the space $%
\widehat{{\bf g}}_{+}^{*}$ and symplectic leaves of this bracket are
coadjoint orbits of the algebra $\widehat{{\bf g}}_{+}$ in the space $%
\widehat{{\bf g}}_{+}^{*}$ . The Lax operators of integrable models lie in
this space. In our realization they have the form (\ref{2.5.1}) with respect
to the spectral parameter $\lambda $ . It is evident from the definition (%
\ref{2.7a}) that the Poisson structure of such models will be connected only
with the poles of  $X\left( \lambda \right) $ in (\ref{2.5.1}).
while local conservation laws are connected with the
asymptotic expansion of the monodromy matrix in the neighbourhood of the poles
of the Lax operator and hence  they depend on the poles of the  function $%
\phi ^{-1}\left( \lambda \right) X\left( \lambda \right) $ in (\ref{2.5.1}).

In the tensor notations introduced above this bracket has the form:

\begin{equation}
\label{2.7b}
\begin{array}{c}
\left\{ L_1\left( x,\lambda \right) ,L_2\left( y,\mu \right) \right\}
=\left[ a,L_1\left( x,\lambda \right) +L_2\left( y,\mu \right) \right]
\delta \left( x-y\right) + \\
+\left[ s,L_1\left( x,\lambda \right) -L_2\left( y,\mu \right) \right]
\delta \left( x-y\right) +2s\delta ^{\prime }\left( x-y\right) , \\
\text{here }a\text{ and }r\text{ are kernels of the operators:} \\ a=\frac
12\left( r-r^{*}\right) ,s=\frac 12\left( r+r^{*}\right) \\
\text{in the scalar product (\ref{2.5}).}
\end{array}
\end{equation}

For instance, $P_{-}$ has the following kernel

\begin{equation}
\label{2.9}P_{-}\left( \lambda ,\mu \right) =\frac t{\lambda -\mu }\phi
\left( \mu \right) ^{-1}.
\end{equation}

Moreover,

\begin{equation}
\label{2.10}s=-P_{-}+\phi ^{-1}P_{-}\phi ,
\end{equation}

so that

\begin{equation}
\label{2.11}s\mid _{{\bf g}_{-}}=0,s\mid _{{\bf g}_{+}}=\phi ^{-1}P_{-}\phi
\mid _{{\bf g}_{+}},
\end{equation}

$s\mid _{{\bf g}_\nu }\neq 0$ if and only if $\phi $ has a pole at the point
$\nu $ .

\begin{equation}
\label{2.12}a=I-P_{-}-\phi ^{-1}P_{-}\phi .
\end{equation}

Thus we have

\begin{equation}
\label{2.13}
\begin{array}{c}
s\left( \lambda ,\mu \right) =\frac t{\lambda -\mu }\left( \phi \left(
\lambda \right) ^{-1}-\phi \left( \mu \right) ^{-1}\right) , \\
a\left( \lambda ,\mu \right) =\frac t{\lambda -\mu }\left( \phi \left(
\lambda \right) ^{-1}-\phi \left( \mu \right) ^{-1}\right) .
\end{array}
\end{equation}

For another point of view on the bracket (\ref{2.7a}) see \cite{rs1}.

Now we are going back to the principal chiral field. For this model there
exists the Zakharov-Mikhailov Lax pair \cite{ft}:

\begin{equation}
\label{2.14}
\begin{array}{c}
L=-\frac 1{1-\lambda ^2}\left( l_x+\lambda l_t\right) , \\
T=-\frac 1{1-\lambda ^2}\left( l_t+\lambda l_x\right) .
\end{array}
\end{equation}

The equations of motion (\ref{1.15}) are expressed as the zero curvature
condition:

\begin{equation}
\label{2.15}\left[ \partial _x-L,\partial _t-T\right] =0.
\end{equation}

We propose the following Lie-algebraic interpretation of this pair. Let us
consider the algebra ${\bf g}_{{\cal A}}$ with the scalar product (\ref{2.5}%
), where

\begin{equation}
\label{2.16}\phi \left( \lambda \right) =2\frac{\lambda ^2-1}{\lambda ^2}.
\end{equation}

. Our main result is the theorem:

\begin{theorem}
The r-matrix Lie-Poisson bracket (\ref{2.7a}) for the Lax operator (\ref
{2.14}) coincides with the brackets (\ref{1.12}) for the canonical Poisson
structure of the principal chiral field if $\phi $ in the definition (\ref
{2.7a}) is given by the formula (\ref{2.16}). In this case we must put in (%
\ref{2.7b}):
\end{theorem}

\begin{equation}
\label{2.17}
\begin{array}{c}
s\left( \lambda ,\mu \right) =-\frac 12
\frac{t\left( \lambda +\mu \right) }{\left( 1-\lambda ^2\right) \left( 1-\mu
^2\right) }, \\ a\left( \lambda ,\mu \right) =\frac 12\frac t{\lambda -\mu
}\left( \frac{\lambda ^2}{1-\lambda ^2}+\frac{\mu ^2}{1-\mu ^2}\right) .
\end{array}
\end{equation}

The theorem is verified by a direct computation. Thus the principal chiral
field is described by the general Lie-algebraic scheme used in the Classical
Inverse Scattering Method \cite{ft},\cite{s1}. We conclude this section with
the formula allowing to restore the square of moment (\ref{1.11}) using only
the monodromy matrix of our model. It should be mentioned that the moment (%
\ref{1.11}) is not well defined quantity on the space ${\em G}\backslash
T^{*}\widehat{{\em G}}$ because we may restore the variable $r_t$ on the
space ${\em G}\backslash T^{*}\widehat{{\em G}}$ only up to constant loops.
The well defined quantity is the square of the moment $\mu _L^{{\em G}}$
with respect to the scalar product $tr$ on ${\bf g}$ .

Let us consider the equation for the monodromy matrix:

\begin{equation}
\label{2.18}\partial _x\Psi \left( x,\lambda \right) =-\frac 1{1-\lambda
^2}\left( l_x+\lambda l_t\right) \Psi \left( x,\lambda \right) ,\Psi \left(
0,\lambda \right) =I,
\end{equation}

so that $M\left( \lambda \right) =\Psi \left( 2\pi ,\lambda \right) $ .We
have:

\begin{equation}
\label{2.19}
\begin{array}{c}
\partial _x\Psi \left( x,0\right) =-l_x\Psi \left( x,0\right)
=-g^{-1}\partial _xg\Psi \left( x,0\right) , \\
\text{so that modulo left actions of constant loops }\Psi \left( x,0\right)
=g^{-1}\left( x\right) .
\end{array}
\end{equation}

It means that there exist an element $g_0\in {\em G}$ such that $\Psi \left(
x,0\right) g_0=g^{-1}\left( x\right) $ .

Let

\begin{equation}
\label{2.20}\stackrel{\bullet }{\Psi }\left( x,0\right) =\frac \partial
{\partial \lambda }\mid _{\lambda =0}\Psi \left( x,\lambda \right) .
\end{equation}

For this function we have the equation:

\begin{equation}
\label{2.21}\partial _x\stackrel{\bullet }{\Psi }\left( x,0\right)
=-l_tg^{-1}\left( x\right) -l_x\stackrel{\bullet }{\Psi }\left( x,0\right) .
\end{equation}

This equation has the following general solution:

\begin{equation}
\label{2.22}\stackrel{\bullet }{\Psi }\left( x,0\right) =-g^{-1}\left(
x\right) \int\limits_0^xg\left( y\right) l_t\left( y\right) g\left( y\right)
^{-1}dy,
\end{equation}

and

\begin{equation}
\label{2.23}\stackrel{\bullet }{M}\left( 0\right) =\frac d{d\lambda }\mid
_{\lambda =0}M\left( \lambda \right) =M\left( 0\right) \int\limits_0^{2\pi
}r_t\left( y\right) dy,
\end{equation}

where we use (\ref{1.11}). Finally we modulo left action of constant loops :

\begin{equation}
\label{2.24}\mu _L^{{\em G}}\left( g,l_t\right) =\int\limits_0^{2\pi
}r_t\left( x\right) dx=M^{-1}\left( 0\right) \stackrel{\bullet }{M}\left(
0\right) .
\end{equation}

We remind that on the phase space of the principal chiral field one must put
$\mu _L^{{\em G}}=0$. Our formula allows to impose this condition only by
means of the additional constraint for the monodromy matrix.

\section{The coadjoint orbits formulation for a nonlinear sigma-model.}

We shall try to generalize the construction of the previous section for a
nonlinear sigma-model. To make this it is natural to consider so-called
twisted  algebra of ad\`eles.

Let $\sigma $ be the involution defined by the Cartan decomposition (\ref
{1.16}):

\begin{equation}
\label{3.1}\sigma :{\bf g}\rightarrow {\bf g},\sigma \mid _{{\bf k}%
}=id,\sigma \mid _{{\bf p}}=-id.
\end{equation}

This involution gives rise to an involution $\widehat{\sigma }$ on the
algebra ${\bf g}_{{\cal A}}$ :

\begin{equation}
\label{3.2}
\begin{array}{c}
\widehat{\sigma }:{\bf g}_{{\cal A}}\rightarrow {\bf g}_{{\cal A}},\widehat{%
\sigma }\left( X\left( \lambda \right) \right) =\sigma X\left( -\lambda
\right) , \\ X\left( \lambda \right) =\bigoplus\limits_{\nu \in \overline{%
{\bf C}}}X_\nu \left( \lambda _\nu \right) \in {\bf g}_{{\cal A}},\sigma
X\left( -\lambda \right) =\bigoplus\limits_{\nu \in {\bf C}}\sigma X_\nu
\left( -\lambda -\nu \right) \oplus \sigma X_\infty \left( \frac 1\lambda
\right) .
\end{array}
\end{equation}

We define the twisted algebra of ad\`eles ${\bf g}_{{\cal A}}^\sigma $ as
the subalgebra of elements of ${\bf g}_{{\cal A}}$ invariant under $\widehat{%
\sigma }$ :

\begin{equation}
\label{3.3}{\bf g}_{{\cal A}}^\sigma =\left\{ X\left( \lambda \right) \in
{\bf g}_{{\cal A}}:\widehat{\sigma }\left( X\left( \lambda \right) \right)
=X\left( \lambda \right) \right\} .
\end{equation}

We introduce the following notations:

\begin{equation}
\label{3.4}{\bf g}_{-}^\sigma ={\bf g}_{{\cal A}}^\sigma \cap {\bf g}_{-},%
{\bf g}_{+}^\sigma ={\bf g}_{{\cal A}}^\sigma \cap {\bf g}_{+}.
\end{equation}

As above, there exists a direct decomposition of the linear space ${\em g}_{%
{\cal A}}^\sigma $ :

\begin{equation}
\label{3.5}{\bf g}_{{\cal A}}^\sigma ={\bf g}_{+}^\sigma \stackrel{.}{+}{\bf %
g}_{-}^\sigma .
\end{equation}

We define on ${\bf g}_{{\cal A}}^\sigma $ the invariant scalar product:

\begin{equation}
\label{3.6}\left( X\left( \lambda \right) ,Y\left( \lambda \right) \right)
=\sum\limits_{\nu \in \overline{{\bf C}}}\stackunder{\nu }{Res}\text{ }%
trX_\nu \left( \lambda _\nu \right) Y_\nu \left( \lambda _\nu \right) \frac{%
d\lambda }\lambda .
\end{equation}

Let $P_{\pm }^\sigma $ be the projectors on ${\bf g}_{\pm }^\sigma $ .Then
we construct the standard r-matrix on ${\bf g}_{{\cal A}}^\sigma $ :

\begin{equation}
\label{3.7}r^\sigma =P_{+}^\sigma -P_{-}^\sigma .
\end{equation}

This r-matrix is not skew-symmetric with respect to the scalar product (\ref
{3.6}), because the subalgebras ${\bf g}_{\pm }^\sigma $ are not isotropic
with respect to this scalar product. The projector $P_{-}^\sigma $ has the
following kernel:

\begin{equation}
\label{3.8}P_{-}^\sigma \left( \lambda ,\mu \right) =t_A\frac{\mu ^2}{%
\lambda ^2-\mu ^2}+t_B\frac{\lambda \mu }{\lambda ^2-\mu ^2},
\end{equation}

and $r^\sigma $ has the symmetric part with the kernel:

\begin{equation}
\label{3.9}s\left( \lambda ,\mu \right) =t_A,
\end{equation}

and the skew-symmetric part:

\begin{equation}
\label{3.10}a\left( \lambda ,\mu \right) =-t_A\frac{\lambda ^2+\mu ^2}{%
\lambda ^2-\mu ^2}-t_B\frac{2\lambda \mu }{\lambda ^2-\mu ^2}.
\end{equation}

With the same purposes as in the previous section, we shall use the deformed
scalar product:

\begin{equation}
\label{3.11}\left( X\left( \lambda \right) ,Y\left( \lambda \right) \right)
_\phi =\sum\limits_{\nu \in \overline{{\bf C}}}\stackunder{\nu }{Res}\text{ }%
trX_\nu \left( \lambda _\nu \right) Y_\nu \left( \lambda _\nu \right) \phi
\left( \lambda ^2\right) \frac{d\lambda }\lambda ,
\end{equation}

here we must use only a scalar rational function $\phi $ depending on the
variable $\lambda ^2$ to define the scalar product correctly.

Then with respect to this scalar product the kernel of the operator $%
P_{-}^\sigma $ has the form:

\begin{equation}
\label{3.14}P_{-}^\sigma \left( \lambda ,\mu \right) _\phi =\left( t_A\frac{%
\mu ^2}{\lambda ^2-\mu ^2}+t_B\frac{\lambda \mu }{\lambda ^2-\mu ^2}\right)
\phi \left( \mu ^2\right) ^{-1}.
\end{equation}

For the description of the Poisson structure of a nonlinear sigma-model we
remind that , as well as for the principal chiral field , there exists a Lax
pair for this model:

\begin{equation}
\label{3.12}
\begin{array}{c}
L=-\left( A_x+\frac \lambda 2\left( B_x+B_t\right) +\frac 1{2\lambda }\left(
B_x-B_t\right) \right) , \\
T=-\left( +\frac \lambda 2\left( B_x+B_t\right) -\frac 1{2\lambda }\left(
B_x-B_t\right) \right) ,
\end{array}
\end{equation}

and the equations of a motion have the form (\ref{2.15}).

We choose the following form of $\phi $ :

\begin{equation}
\label{3.13}\phi \left( \lambda ^2\right) =-\frac{4\lambda ^2}{\left(
\lambda ^2-1\right) ^2}.
\end{equation}

For our purposes we need to deform the bracket (\ref{2.7a}). Let us define
the operator $r_0$ acting on the algebra ${\bf g}_{{\cal A}}^\sigma $ :

\begin{equation}
\label{3.15}r_0\left( X\left( \lambda \right) \right) =\left( P_{{\bf k}%
}r^\sigma X\right) \left( 1\right) .
\end{equation}

This operator has the kernel:

\begin{equation}
\label{3.16}P_{-}^{{\bf k}}\left( \mu \right) =t_A\frac{\mu ^2}{1-\mu ^2}%
\phi \left( \mu ^2\right) ^{-1}
\end{equation}

The operator

\begin{equation}
\label{3.17}\widehat{r}=r^\sigma -r_0
\end{equation}

does not satisfy the classical Yang-Baxter equation, so that this operators
may not be used as an r-matrix for the definition of the Poisson brackets (%
\ref{2.7a}). But we may use this operator for the definition of the Poisson
brackets (\ref{2.7a}) of the functions $\varphi $ for which the derivatives $%
X_\varphi $ lie in the kernel of the operator $r_0$ , because the
Lie-Poisson bracket (\ref{2.7a}) depends on $r$ only via combinations $%
\widehat{r}X_\varphi $. For instance it is not difficult to verify that
derivatives of functions depending on the coefficients of the Lax operator (%
\ref{3.12}) satisfy this condition.

Hence we may define the Poisson brackets of the Lax operator (\ref{3.12}) by
the formula (\ref{2.7a}) using the operator $\widehat{r}$ . Our main result
is

\begin{theorem}
Let $\phi $ be given by the formula (\ref{3.13}); choose the r-matrix $%
\widehat{r}$ as in (\ref{3.17}) .Then for the Lax operator $L$ (\ref{3.12})
the r- bracket (\ref{2.7a}) coincides with the canonical Poisson structure
of the nonlinear sigma-model (\ref{1.19}). In this case we must put in (\ref
{2.7b}):
\end{theorem}

\begin{equation}
\label{3.21}
\begin{array}{c}
s\left( \lambda ,\mu \right) =
\frac{t_B}4\left( \frac 1{\lambda \mu }-\lambda \mu \right) , \\ a\left(
\lambda ,\mu \right) =P_{-}^\sigma \left( \mu ,\lambda \right) _\phi +P_{-}^{%
{\bf k}}\left( \lambda \right) -P_{-}^\sigma \left( \lambda ,\mu \right)
_\phi -P_{-}^{{\bf k}}\left( \mu \right) .
\end{array}
\end{equation}
{\bf \ }

We  conclude this section with a formula for the square of moment map (\ref
{1.11}). Without comments we present the straightforward calculation which
is similar to the one in the previous section (\ref{2.18})-(\ref{2.24})

$$
\begin{array}{c}
\partial _x\Psi \left( x,\lambda \right) =-\left( A_x+\frac \lambda 2\left(
B_x+B_t\right) +\frac 1{2\lambda }\left( B_x-B_t\right) \right) \Psi \left(
x,\lambda \right) ,\Psi \left( 0,\lambda \right) =I, \\
M\left( \lambda \right) =\Psi \left( 2\pi ,\lambda \right) , \\
\partial _x\Psi \left( x,1\right) =-\left( A_x+B_t\right) \Psi \left(
x,1\right) =-l_x\Psi \left( x,1\right) ,
\end{array}
$$

$$
\begin{array}{c}
\Psi \left( x,1\right) =g^{-1}\left( x\right) ,
\text{modulo left action of constant loops,} \\ \stackrel{\bullet }{\Psi }%
\left( x,1\right) =\frac \partial {\partial \lambda }\mid _{\lambda =1}\Psi
\left( x,\lambda \right) , \\ \partial _x\stackrel{\bullet }{\Psi }\left(
x,1\right) =-l_x\stackrel{\bullet }{\Psi }\left( x,1\right) -B_t\Psi \left(
x,1\right) =
\end{array}
$$

$$
\begin{array}{c}
=-l_x
\stackrel{\bullet }{\Psi }\left( x,1\right) -B_tg^{-1}\left( x\right) , \\
\stackrel{\bullet }{\Psi }\left( x,1\right) =-g^{-1}\left( x\right)
\int\limits_0^xg\left( y\right) B_t\left( y\right) g^{-1}\left( y\right)
dy=\Psi \left( x,1\right) \int\limits_0^xr_t\left( y\right) dy, \\
M^{-1}\left( 1\right) \stackrel{\bullet }{M}\left( 1\right)
=\int\limits_0^{2\pi }r\left( x\right) dx=\mu _L^{{\em G}}\left( g,l\right)
\text{ modulo constant loops.}
\end{array}
$$

On the phase space of the chiral field
$\mu _L^{{\em G}}\left( g,l\right) =0$ . Our formula for the moment allows to
impose this constraint in explicit form as an additional condition on the
monodromy matrix.

\end{document}